\documentclass[amsmath,amssymb]{jpsj3} 
\usepackage{color}
\usepackage{graphicx}
\usepackage{dcolumn}
\usepackage{bm}
\usepackage{booktabs}
\DeclareFontFamily{U}{rsfs}{\skewchar\font127}
\DeclareFontShape{U}{rsfs}{m}{n}{<-6>rsfs5<6-8.5>rsfs7<8.5->rsfs10}{}
\DeclareSymbolFont{rsfs}{U}{rsfs}{m}{n}
\DeclareSymbolFontAlphabet{\mathrsfs}{rsfs}
\DeclareRobustCommand*\rsfs{\@fontswitch\relax\mathrsfs}

\title{
Impact of Local Lattice Disorder on Spin and Orbital Orders in Ca$_{2-x}$Sr$_x$RuO$_4$
}

\author{Takuya Sugimoto$^1$, Daiki Ootsuki$^2$, Takashi Mizokawa$^{1,2}$}
\inst{
$^1$Department of Complexity Science and Engineering, University of Tokyo, 5-1-5 Kashiwanoha, Kashiwa, 
Chiba 277-8561, Japan\\
$^2$Department of Physics, University of Tokyo, 5-1-5 Kashiwanoha, Kashiwa, 
Chiba 277-8561, Japan\\
}

\date{\today}

\abst{
We have studied relationship between local lattice disorder, Ru 4$d$ spin-orbit interaction,
and global spin/orbital orders of Ca$_{2-x}$Sr$_x$RuO$_4$ in the Sr-rich and Ca-rich regions 
of its phase diagram by using unrestricted Hartree-Fock calculation 
on a $8\times8$ RuO$_4$ lattice model.
The calculations show that the local elongation of RuO$_6$ octahedron 
in an antiferromagnetic insulator Ca$_2$RuO$_4$ induces local orbital change 
with making the Mott gap narrower. 
On the other hand, local compression of RuO$_6$ octahedron in a paramagnetic 
metal Sr$_2$RuO$_4$ induces global antiferromagnetic or ferromagnetic states. 
This result is consistent with a recent systematic $\mu$SR study by Carlo {\it et al.} 
which has revealed the static antiferromagnetic order at low temperature 
in the Sr-rich region.
}

\kword{layered ruthenates, antiferromagnetism, orbital degeneracy, Mott insulator}

\begin{document}
\maketitle

\newpage

\section{Introduction}
The layered perovskite Ca$_{2-x}$Sr$_x$RuO$_4$ (CSRO) system has been attracting 
considerable interest due to the interesting evolution from the spin-triplet
superconducting state in Sr$_2$RuO$_4$ to the Mott insulating state 
in Ca$_2$RuO$_4$ \cite{Maeno94,Nakatsuji00a,Nakatsuji00b,Imada98}. 
The structural phase diagram of CSRO exhibits an interesting interplay 
between titling, rotation, and Jahn-Teller distortion of RuO$_6$ 
octahedron \cite{Friedt01}. The magnetic and electronic properties of CSRO 
strongly correlate with the structural distortions. The Mott transition 
at $x = 0.0$ (Ca$_2$RuO$_4$) is accompanied by the Ru 4$d$ orbital change 
due to the Jahn-Teller distortion \cite{Mizokawa01,Mizokawa04}. 
For $x \leq 0.2$, CSRO is an antiferromagnetic insulator at low temperature 
due to the Jahn-Teller driven compression of RuO$_6$ octahedron along the $c$-axis.
For $0.2 \leq x \leq 0.5$, the tilting of RuO$_6$ octahedron provides 
orthorhombic distortion, and the magnetic susceptibility shows a heavy 
Fermion behavior at low temperature. The orthorhombic distortion seems to 
suppress the ferromagnetic and/or small-$q$ antiferromagnetic fluctuation 
while it still enhances the mass renormalization towards $x= 0.2$. 
Although the mass enhancement around $x= 0.2$ is claimed to be explained 
by orbital selective Mott transition \cite{Anisimov02}, existence of 
orbital selective Mott transition in multi-band Hubbard models depends on 
the details of parameters in the Hubbard Hamiltonians \cite{Liebsch03, Koga04}.
It is still controversial whether the orbital selective Mott transition 
of the multi-band Ru 4$d$ electrons is relevant for the electronic phase diagram 
of CSRO or not. As for the tiny Sr substitution in the Mott insulating state 
of Ca$_2$RuO$_4$, the transport properties are dramatically changed 
by the Sr doping \cite{Nakatsuji04}. 

Very recently, a systematic $\mu$SR study has revealed that static 
antiferromagnetic order exists at low temperature even in the Sr-rich 
region ($1.5 \leq x \leq 2.0$) of its phase diagram \cite{Carlo12}.
This indicates that the local distortion of RuO$_6$ octahedron introduced
by the Ca substitution plays important roles to stabilize 
the antiferromagnetic state. 
On the other hand, in the Ca-rich region ($0.0 \leq x \leq 0.5$), 
the Sr-substitution reduces the magnitude of the Jahn-Teller distortion,
tilting, and rotation of the RuO$_6$ octahedron in Ca$_2$RuO$_4$ and, consequently,
destroys the spin and orbital orders of Ca$_2$RuO$_4$.
In order to gain deeper understandings of the phase diagram, 
it is important to study the Ru 4$d$ spin-orbital states 
using a realistic model in which the effects of spin-orbit interaction 
and lattice distortions are considered. 

In this paper, we investigate the Sr or Ca doping effects on the electronic 
structure of CSRO system at both ends of its phase diagram (which are 
Sr$_2$RuO$_4$ and Ca$_2$RuO$_4$) by means of unrestricted Hartree-Fock (HF) 
calculation, which includes the spin-orbit interaction and lattice distortion 
induced by the chemical substitution.

\section{Method of calculation}

We use the multiband $d$-$p$ model where full degeneracy of the Ru $4d$ orbitals 
and the O $2p$ orbitals are taken into account. The Hamiltonian is given by
\begin{align}
\hat{\mathrsfs{H}} =& \hat{\mathrsfs{H}}_p + \hat{\mathrsfs{H}}_d + \hat{\mathrsfs{H}}_{pd}  \notag \\
 \hat{\mathrsfs{H}}_p =& \sum_{kl\sigma} \epsilon^p_k p^{\dagger}_{kl\sigma} p_{kl\sigma} + \sum_{kll'\sigma}V^{pp}_{kll'}p^{\dagger}_{kl\sigma} p_{kl'\sigma} + \text{h.c.} \notag \\
\hat{\mathrsfs{H}}_d =& \epsilon^0_d \sum_{i \alpha m \sigma} d^{\dagger}_{i \alpha m \sigma}d_{i \alpha m \sigma} + \sum_{i \alpha mm'\sigma \sigma'}h_{mm'\sigma \sigma'}d^{\dagger}_{i \alpha m \sigma}d_{i \alpha m' \sigma'}    \notag \\
&+ u\sum_{i \alpha m} d^{\dagger}_{i \alpha m \uparrow}d_{i \alpha m \uparrow} d^{\dagger}_{i \alpha m \downarrow}d_{i \alpha m \downarrow}  \notag \\
& +u'\sum_{i \alpha m m'} d^{\dagger}_{i \alpha m \uparrow}d_{i \alpha m \uparrow}  d^{\dagger}_{i \alpha m \downarrow}d_{i \alpha m \downarrow} \notag \\
&+ (u'-j)\sum_{i \alpha mm'\sigma}  d^{\dagger}_{i \alpha m \sigma}d_{i \alpha m \sigma}  d^{\dagger}_{i \alpha m' \sigma}d_{i \alpha m' \sigma}  \notag \\
&+ j \sum_{i \alpha mm'} d^{\dagger}_{i \alpha m \uparrow} d_{i \alpha m' \uparrow} d^{\dagger}_{i \alpha m' \downarrow}d_{i \alpha m \downarrow}  \notag \\
&+ j' \sum_{i \alpha mm'} d^{\dagger}_{i \alpha m \uparrow} d_{i \alpha m' \uparrow}  d^{\dagger}_{i \alpha m \downarrow}d_{i \alpha m' \downarrow} \notag \\
\hat{\mathrsfs{H}}_{pd} =& \sum_{kml\sigma}V^{pd}_{kml}d^{\dagger}_{km\sigma} p_{kl\sigma} + \text{h.c.} \notag 
\end{align}

\begin{figure}
\includegraphics[width=0.9\textwidth]{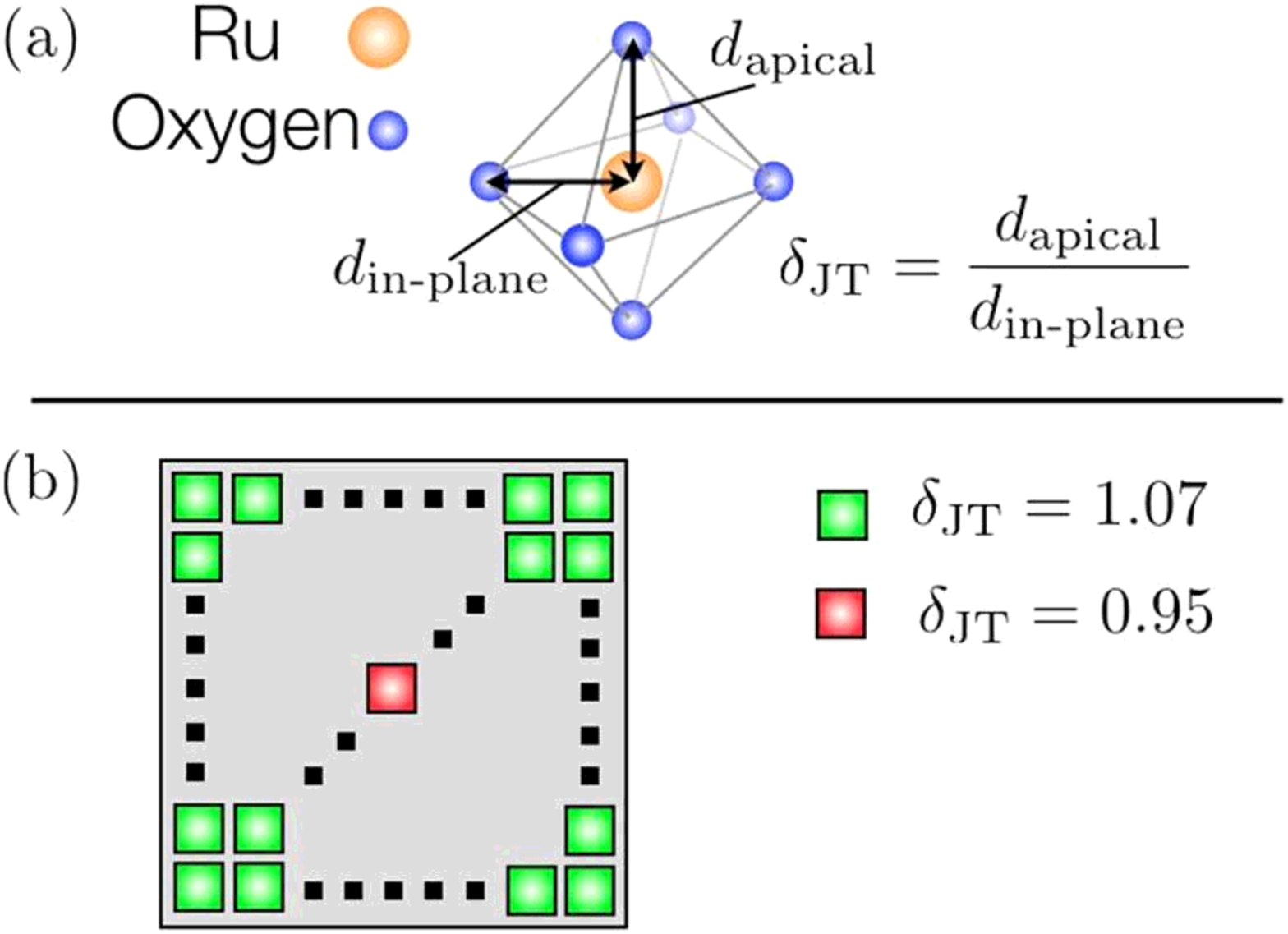}
\label{fig:deltaJT}
\end{figure}

Here, $d^{\dagger}_{i \alpha m \sigma}$ are creation operators for the Ru $4d$ electrons 
at site $\alpha$ of the $i^{\text{th}}$ unit cell and $d^{\dagger}_{km\sigma}$ and $p^{\dagger}_{kl\sigma}$ 
are creation operators for Bloch electrons which are constructed from the $m^{\text{th}}$ component 
of the $4d$ orbitals and from the $l^{\text{th}}$ component of the O $2p$ orbitals, respectively, 
with wave vector $\bm{k}$. The matrix $h_{mm'\sigma \sigma'}$ denotes the spin-orbit interaction 
and the effects of crystal field splitting. The magnitude of the spin-orbit interaction 
for the Ru $4d$ orbital is fixed as 0.15 eV. The transfer integrals between the O $2p$ orbitals 
$V^{pp}_{kll'}$ are given by Slater-Koster parameters $(pp\sigma)$ and $(pp\pi)$ which are fixed 
at $0.60$ eV and $-0.15$ eV respectively. The transfer integrals between the Ru $4d$ and O $2p$ orbitals
 $V^{pd}_{kml}$ are represented by $(pd\pi)$ and $(pd\sigma)$. They are fixed as $(pd\sigma) = -2.8$ eV 
and $(pd\pi) = 1.26$ eV for the longer in-plane Ru-O bond of Ca$_2$RuO$_4$ whereas $(pd\sigma) = -3.4$ eV 
and $(pd\pi) = 1.53$ eV for the shorter in-plane Ru-O bond of Sr$_2$RuO$_4$. The summary of
the material-dependent parameters are shown in Table \ref{parameters}.
The tilting of the RuO$_6$ octahedron is included for Ca$_2$RuO$_4$. 
The distortion parameter $\delta_{\text{JT}}$ is defined as 
$\delta_{\text{JT}} = d_{\text{apical}}/d_{\text{in-plane}}$
which is the ratio between the apical and in-plane Ru-O bond distances.
The distortion parameter $\delta_{\text{JT}}$ is utilized to express the elongation/compression 
of RuO$_6$ octahedron as in Fig. \ref{fig:deltaJT}(a). 
In the Sr-rich (Ca-rich) region, $\delta_{\text{JT}} = 1.07$ (0.95) for the host lattice and 
$\delta_{\text{JT}} = 0.95$ (1.07) for the locally distorted site. 
When the RuO$_6$ octahedron is distorted, the transfer integrals are scaled by Harrison's rule. 
The intra-atomic Coulomb interactions between Ru $4d$ electrons are given by Kanamori parameters. 
They are fixed as $u = u'+ j + j' = 3.0$ eV and $j = j' = 0.5$ eV. The charge-transfer energy $\Delta$ 
(fixed as $-0.4$ eV) is defined by $\epsilon_d -\epsilon_p +4U$, where $\epsilon_d$ and $\epsilon_p$ 
are the energies of the bare Ru $4d$ and O $2p$ orbitals and $U [=u-(20/9)j]$ is the multiplet-averaged 
$d$-$d$ Coulomb interaction fixed at 1.89 eV. 

We set the $8\times 8$ supercell with periodic boundary conditions and 
put the Ru $4d$ and O $2p$ electrons on its each site. The local distortion 
is introduced at one Ru site of the supercell as shown in Figure \ref{fig:deltaJT}(b). 
The total number of electron in the supercell is 1792 
(256 Ru $4d$ and 1536 O $2p$ electrons). 
The HF mean-field treatment is applied to the two-body part in $\hat{\mathrsfs{H}}_d$ 
by replacing its average values, for instance,

\begin{align}
u \sum_{i \alpha m} & d^{\dagger}_{i \alpha m \uparrow}d_{i \alpha m \uparrow} d^{\dagger}_{i \alpha m \downarrow} d_{i \alpha m \downarrow} \notag \\
\rightarrow &\notag \\
 & u\sum_{i \alpha m}\langle d^{\dagger}_{i \alpha m \uparrow}d_{i \alpha m \uparrow} \rangle d^{\dagger}_{i \alpha m \downarrow}d_{i \alpha m \downarrow}  \notag \\
 &+ u\sum_{i \alpha m} d^{\dagger}_{i \alpha m \uparrow}d_{i \alpha m \uparrow} \langle d^{\dagger}_{i \alpha m \downarrow}d_{i \alpha m \downarrow}\rangle  \notag \\
&-  u\sum_{i \alpha m} \langle d^{\dagger}_{i \alpha m \uparrow}d_{i \alpha m \uparrow} \rangle \langle d^{\dagger}_{i \alpha m \downarrow}d_{i \alpha m \downarrow}\rangle \notag
\end{align}

In this Hartree-Fock calculation, we input the initial values of the order parameters such as
$\langle d^{\dagger}d \rangle$ and diagonalize the mean-field Hamiltonian to get a set of eigen functions. 
Then the order parameters can be calculated using the obtained eigen functions. This self-consistency cycle
 is iterated until the successive difference of all the order parameters converge less than $10^{-4}$.

\section{Results and Discussion}

We calculate the following four cases to see the doping effects. For the Ca-rich region,
we calculate the orbital population of Ca$_2$RuO$_4$ and Ca$_{2-x}$Sr$_{x}$RuO$_4$ 
with a single-site distortion by the tiny Sr doping (let us label these two 
cases as case (i) and (ii), respectively).
For the Sr-rich region, we calculate Sr$_2$RuO$_4$ and Ca$_{x}$Sr$_{2-x}$RuO$_4$ 
with a single-site distortion by the Ca doping (let us label these two cases as case (iii) 
and (iv), respectively). 
Since we set the $8\times8$ supercell, the ratio of one-site transposition is equal
to $1/64 \simeq 0.0156$. In terms of the doping amount in this CSRO system, $1.56\%$
is equivalent to $x=0.0312$ in Ca$_{2-x}$Sr$_{x}$RuO$_4$ and Ca$_{x}$Sr$_{2-x}$RuO$_4$, 
respectively.

\subsection{Ca-rich region}

The present HF calculation using the reasonable parameter set 
predicts that Ca$_2$RuO$_4$ is an antiferromagnetic insulator. 
The compression of the RuO$_6$ octahedra stabilizes the Ru 4$d$ $xy$ orbital
compared to the Ru 4$d$ $yz/zx$ orbitals (Jahn-Teller type energy splitting). 
The combination of the Jahn-Teller type energy splitting and the Ru 4$d$ spin-orbit 
interaction can provide an interesting magnetic anisotropy. The antiferromagnetic
insulating state with the in-plane ($x$ or $y$ axis) spin direction is lower in energy
than that with the out-of-plane ($z$ axis) spin direction. The energy difference
is 37 meV per Ru site for the present parameter set.
The expectation value of $d^{\dagger}d$ for each orbital and spin component
of the Ru $4d$ $t_{2g}$ states for Ca$_2$RuO$_4$ [case (i)] is shown in 
Table \ref{tab:orbpop2} for the in-plane and out-of-plane
antiferromagnetic states. In the out-of-plane case, the complex orbitals
with type of $yz \pm izx$ are unoccupied [namely, occupied by the two Ru 4$d$ $t_{2g}$ 
holes of the Ru$^{4+}$ ($d^4$) configuration] while the $xy$ orbitals are
almost fully occupied by the Ru 4$d$ $t_{2g}$ electrons. Consequently the orbital 
angular momentum along the $z$-axis is formed to align the spin moment along 
the $z$-axis. As for the in-plane case, the complex orbitals with type of
$xy \pm izx$ or $xy \pm iyz$ are unoccupied to give orbital angular momentum 
in the $xy$-plane. 

\begin{figure}
\includegraphics[width=0.9\textwidth]{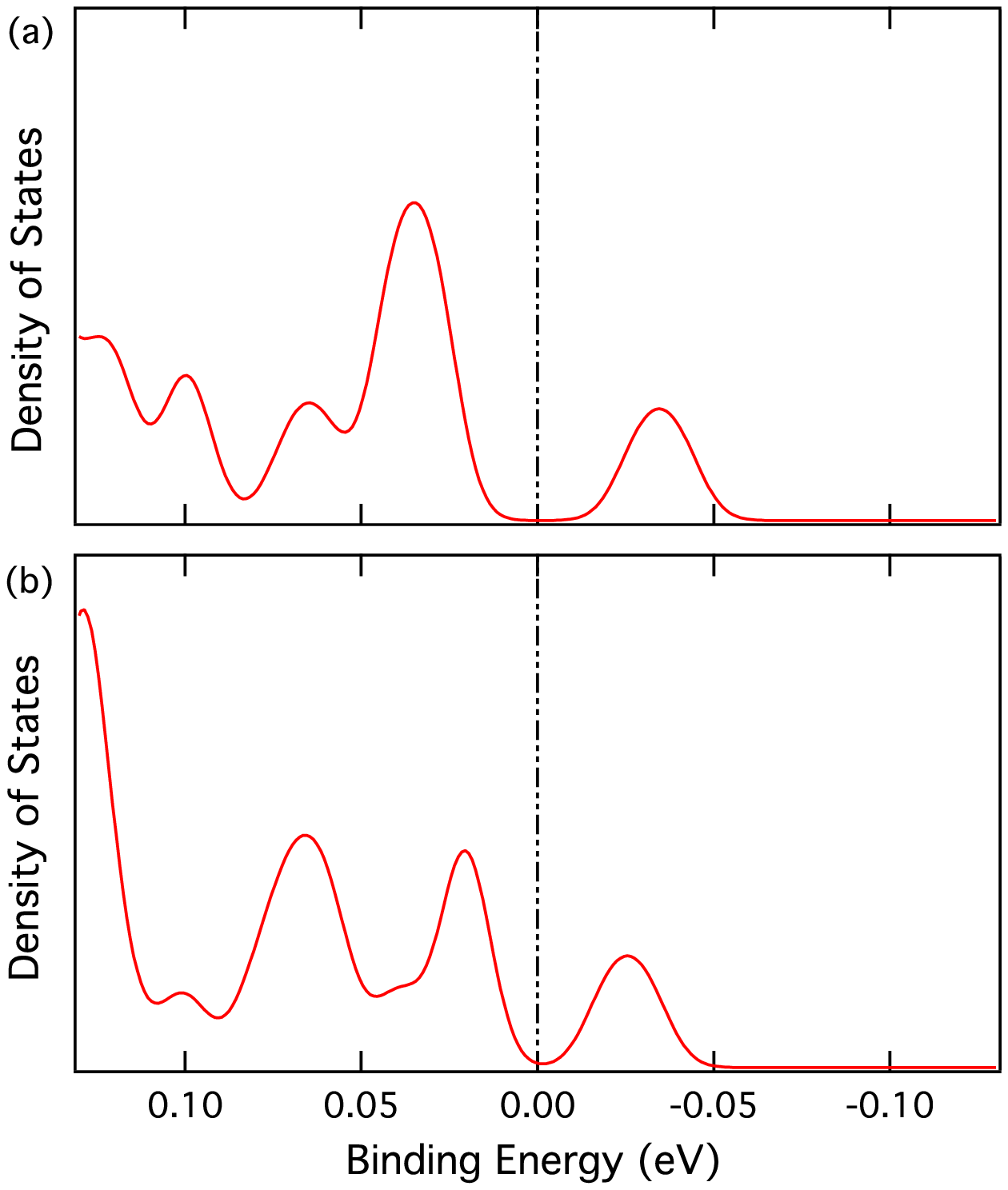}
\label{fig:CaRich}
\end{figure}

Assuming that one of the RuO$_6$ octahedra is elongated due to the tiny Sr doping
in the case (ii), the orbital population of the elongated site would be 
affected due to the reverse of the Jahn-Teller energy splitting 
between the $xy$ and $yz/zx$ orbitals. Interestingly, the orbital population
of the elongated site is dramatically changed for the in-plane antiferromagnetic
state whereas the impact of the elongation is limited for the out-of-plane 
antiferromagnetic state. As shown in Table \ref{tab:orbpop2}, the orbital
population of the elongated site is almost the same as that of the compressed
sites for the out-of-plane antiferromagnetic state. On the other hand, 
the $xy$ orbitals accommodate more holes at the elongated site for the
in-plane antiferromagnetic state. The impact of local reverse of the 
Jahn-Teller energy splitting strongly depends on the global spin direction
due to the strong Ru 4$d$ spin-orbit coupling.
Since the in-plane antiferromagnetic state is realized in Ca$_2$RuO$_4$, 
the present calculation indicates that the local elongation of the RuO$_6$ 
octahedron produces the Ru 4$d$ $xy$ hole which can affect the global spin 
and orbital orders of Ca$_2$RuO$_4$.
The density of states (DOS) calculated for the cases (i) and (ii) 
(which are Ca$_2$RuO$_4$ and Ca$_{2-x}$Sr$_{x}$RuO$_4$ at $x=0.0312$) 
with the in-plane antiferromagnetic states are shown in Figs. \ref{fig:CaRich}(a) and (b), respectively. 
One can see that the Mott gap of Ca$_2$RuO$_4$ becomes narrower 
by the local elongation of RuO$_6$ octahedron ($\delta_{\text{JT}}=1.07$) 
induced by the tiny Sr doping. The reduction of the Mott gap is related to
the local orbital change induced by the reverse of the Jahn-Teller
energy splitting.

\subsection{Sr-rich region}

\begin{figure}
\includegraphics[width=0.9\textwidth]{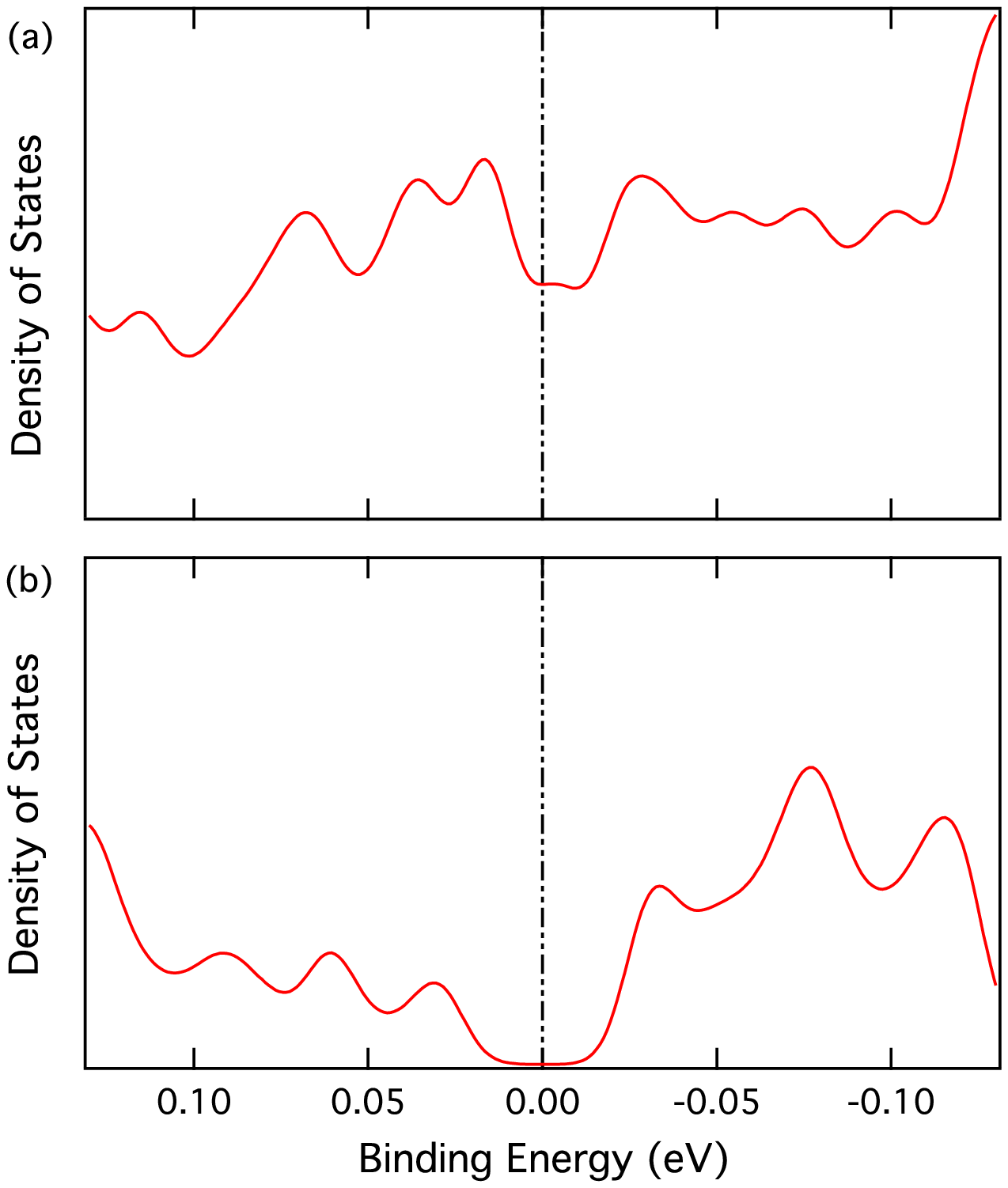}
\label{fig:SrRich}
\end{figure}

The present HF calculation using the reasonable parameter set 
predicts that Sr$_2$RuO$_4$ is a paramagnetic or ferromagnetic metal.
For the present parameter set, the ferromagnetic metallic state is slightly 
lower in energy than the paramagnetic metallic state, and 
the energy difference is about 0.4 meV per Ru site.
In the HF approximation, stability of ferromagnetic or antiferromagnetic 
states tends to be overestimated. Therefore, the present calculation indicates
that the parameter set is reasonable to analyze the paramagnetic metallic state
of Sr$_2$RuO$_4$. 
The expectation value of $d^{\dagger}d$ for each orbital and spin component
of the Ru $4d$ $t_{2g}$ states for Sr$_2$RuO$_4$ [case (iii)] is
shown in Table \ref{tab:orbpop} for the paramagnetic metallic state
and the ferromagnetic metallic state.
In the paramagnetic state, the spin-up and spin-down states have 
the same population for the $xy$, $yx$, and $zx$ orbitals, respectively. 
In the ferromagnetic state, the spin polarization for the $yz/zx$ orbitals
is larger than that of the $xy$ orbital, indicating that the $yz/zx$ states
stabilized by the elongation of the RuO$_6$ octahedra are playing important
roles to provide the ferromagnetic interaction.   
As for the effect of tiny Ca doping to Sr$_2$RuO$_4$ [case (iv)], 
it is assumed that one of the RuO$_6$ octahedra is compressed due to 
the Ca doping. The local compression of the RuO$_6$ octahedron
changes the stability of the paramagnetic state and the nature of the
magnetic state.
The paramagnetic metallic state becomes unstable and
only the ferromagnetic and/or antiferromagnetic states 
are obtained as stable solutions.
The $xy$ orbitals
are more occupied at the compressed site to induce antiferromagnetic 
superexchange interaction. The global antiferromagnetic state is 
stabilized by the local orbital change. The spin direction is
intermediate between the in-plane and out-of-plane directions
probably due to the competition between the Ru 4$d$ spin-orbit interaction 
and the local Jahn-Teller splitting of the Ru 4$d$ levels.
In the stable magnetic state, the in-plane components of the Ru 4$d$ spins
are ferromagnetically aligned whereas the out-of-plane components are
antiferromagnetically arranged.
The DOS calculated for the cases (iii) and (iv) (which are Sr$_2$RuO$_4$ and 
Ca$_{x}$Sr$_{2-x}$RuO$_4$ at $x=0.0312$) are shown in Figs. \ref{fig:SrRich}(a) and (b), respectively. 
The local compression of RuO$_6$ octahedron by the tiny Ca doping induces 
the small band gap at the Fermi level, which can be assigned to the magnetic ordering.

\section{Conclusion}

Investigating the electronic structure of CSRO by unrestricted HF calculation, 
we find the effects of local distortions on spin and orbital orders of the Ru $4d$ $t_{2g}$ states. 
As for the Ca-rich region, we find that the antiferromagnetism survives 
when the single-site distortion in Ca$_2$RuO$_4$ does carry on. 
The orbital state is locally disturbed for the in-plane antiferromagnetic state
resulting in the reduction of the Mott gap.
As for the Sr-rich region, on the other hand, the single-site distortion 
in Sr$_2$RuO$_4$ changes the entire system to the ferromagnetic/antiferromagnetic state from the paramagnetic state. 
The present study captures the trend of effects on Sr$_2$RuO$_4$ 
by the tiny amount of Ca doping, which could explain the recent $\mu$SR study
\cite{Carlo12}. 

\section*{Acknowledgements}
The authors would like to thank Profs. Y. J. Uemura and N. L. Saini 
for fruitful discussions.

\clearpage
Tables:\\
\\
\begin{table}[h]
\caption{Parameter sets for Sr$_2$RuO$_4$ and Ca$_2$RuO$_4$.}
\label{parameters}
\begin{tabular}{cccc}\hline\hline
 \;\;\;\;\;\;\;\;\;\;\;\;\;\;\;\;\;\;\;\; &  \;\;\;($pd\sigma$) for in-plane Ru-O bond\;\;\; & \;\;\;tilting angle \;\;\; & \;\;\;$\delta_{\text{JT}}\;\;\;$ \\ \hline
 Sr$_2$RuO$_4$ & $-3.4$ eV & 0 deg.&  1.07 \\
 Ca$_2$RuO$_4$ & $-2.8$ eV & 12.5 deg. &  0.95 \\ \hline
 \end{tabular}
\end{table}
\\
\\
\begin{table}[h]
\caption{Expectation values of $<d^{\dagger}_{m\sigma}d_{m\sigma}>$ with 
$m=xy, yz, zx$ and $\sigma =\uparrow, \downarrow$ in the Ru $4d$ $t_{2g}$ states
for case (i) and case (ii).}
\label{tab:orbpop2}
\begin{tabular}{c|ccccccc}\hline \hline
case (i) Ca$_2$RuO$_4$ & \; \; & \;\; & \;\; & \;\; & \;\; & \;\; \\ \hline
in-plane ($xy$) AFM & \; $xy\uparrow$ \; & \; $xy\downarrow$ \; & \; $yz\uparrow$ \; & \; $yz\downarrow$ \; & \; $zx\uparrow$ \; & \; $zx\downarrow$ \; \\ 
\;\;\; & 0.95 & 0.95 & 0.53 & 0.76 & 0.52 & 0.77 \\ \hline
out-of-plane ($z$) AFM & \; $xy\uparrow$ \; & \; $xy\downarrow$ \; & \; $yz\uparrow$ \; & \; $yz\downarrow$ \; & \; $zx\uparrow$ \; & \; $zx\downarrow$ \; \\ 
\;\;\; & 0.98 & 0.98 & 0.98 & 0.28 & 0.99 & 0.26 \\ \hline \hline

case (ii) Ca$_{2-x}$Sr$_x$RuO$_4$ & \; \; & \;\; & \;\; & \;\; & \;\; & \;\; \\ \hline
in-plane ($xy$) AFM & \; $xy\uparrow$ \; & \; $xy\downarrow$ \; & \; $yz\uparrow$ \; & \; $yz\downarrow$ \; & \; $zx\uparrow$ \; & \; $zx\downarrow$ \; \\ 
compressed site (Ca site)& 0.97 & 0.96 & 0.64 & 0.64 & 0.64 & 0.64 \\ \hline
in-plane ($xy$) AFM & \; $xy\uparrow$ \; & \; $xy\downarrow$ \; & \; $yz\uparrow$ \; & \; $yz\downarrow$ \; & \; $zx\uparrow$ \; & \; $zx\downarrow$ \; \\ 
elongated site (Sr site)& 0.66 & 0.66 & 0.64 & 0.64 & 0.96 & 0.96 \\ \hline

out-of-plane ($z$) AFM & \; $xy\uparrow$ \; & \; $xy\downarrow$ \; & \; $yz\uparrow$ \; & \; $yz\downarrow$ \; & \; $zx\uparrow$ \; & \; $zx\downarrow$ \; \\ 
compressed site (Ca site)& 0.98 & 0.98 & 0.98 & 0.28 & 0.99 & 0.26 \\ \hline 
out-of-plane ($z$) AFM & \; $xy\uparrow$ \; & \; $xy\downarrow$ \; & \; $yz\uparrow$ \; & \; $yz\downarrow$ \; & \; $zx\uparrow$ \; & \; $zx\downarrow$ \; \\ 
elongated site (Sr site)& 0.99 & 0.98 & 0.99 & 0.28 & 0.99 & 0.26 \\ \hline \hline

\end{tabular}
\end{table}
\\
\\
\begin{table}
\caption{Expectation values of $<d^{\dagger}_{m\sigma}d_{m\sigma}>$ with 
$m=xy, yz, zx$ and $\sigma =\uparrow, \downarrow$ in the Ru $4d$ $t_{2g}$ states
for case (iii) and case (iv).}
\label{tab:orbpop}
\begin{tabular}{c|ccccccc}\hline \hline
case (iii) Sr$_2$RuO$_4$ & \; \; & \;\; & \;\; & \;\; & \;\; & \;\; \\ \hline
PM & \; $xy\uparrow$ \; & \; $xy\downarrow$ \; & \; $yz\uparrow$ \; & \; $yz\downarrow$ \; & \; $zx\uparrow$ \; & \; $zx\downarrow$ \; \\ 
\;\;\; & 0.71 & 0.71 & 0.88 & 0.88 & 0.88 & 0.88 \\ \hline
FM & \; $xy\uparrow$ \; & \; $xy\downarrow$ \; & \; $yz\uparrow$ \; & \; $yz\downarrow$ \; & \; $zx\uparrow$ \; & \; $zx\downarrow$ \; \\ 
\;\;\; & 0.67 & 0.77 & 0.80 & 0.94 & 0.80 & 0.94 \\ \hline \hline

case (iv) Ca$_x$Sr$_{2-x}$RuO$_4$ (Ca-doped) & \; \; & \;\; & \;\; & \;\; & \;\; & \;\; \\ \hline
 AFM & \; $xy\uparrow$ \; & \; $xy\downarrow$ \; & \; $yz\uparrow$ \; & \; $yz\downarrow$ \; & \; $zx\uparrow$ \; & \; $zx\downarrow$ \; \\ 
elongated site (Sr site)& 0.60 & 0.85 & 0.85 & 0.88 & 0.86 & 0.87 \\ \hline
 AFM & \; $xy\uparrow$ \; & \; $xy\downarrow$ \; & \; $yz\uparrow$ \; & \; $yz\downarrow$ \; & \; $zx\uparrow$ \; & \; $zx\downarrow$ \; \\ 
compressed site (Ca site)& 0.79 & 0.94 & 0.61 & 0.89 & 0.61 & 0.89 \\ \hline

FM & \; $xy\uparrow$ \; & \; $xy\downarrow$ \; & \; $yz\uparrow$ \; & \; $yz\downarrow$ \; & \; $zx\uparrow$ \; & \; $zx\downarrow$ \; \\ 
elongated site (Sr site)& 0.67 & 0.78 & 0.80 & 0.94 & 0.80 & 0.94 \\ \hline 
FM & \; $xy\uparrow$ \; & \; $xy\downarrow$ \; & \; $yz\uparrow$ \; & \; $yz\downarrow$ \; & \; $zx\uparrow$ \; & \; $zx\downarrow$ \; \\ 
compressed site (Ca site)& 0.88 & 0.91 & 0.68 & 0.81 & 0.68 & 0.81 \\ \hline \hline

\end{tabular}
\end{table}

\clearpage
Figure captions: \\
Figure 1:\\
(Color online) Schematic diagram of (a) RuO$_6$ octahedron 
with the definition of $\delta_{\text{JT}}$ and (b) the supercell
with a single-site distortion.
\\
\\
Figure 2:\\
(Color online) DOS (calculated with the Ru 4$d$ spin-orbit interaction) 
of (a) Ca$_2$RuO$_4$ and (b) Ca$_{2-x}$Sr$_{x}$RuO$_4$ at $x=0.0312$. 
\\
\\
Figure 3:\\
(Color online) DOS (calculated with the Ru 4$d$ spin-orbit interaction) 
of (a) Sr$_2$RuO$_4$ and (b) Ca$_{x}$Sr$_{2-x}$RuO$_4$ at $x=0.0312$.

\clearpage

\clearpage

\clearpage

\end{document}